\documentclass[doublecol]{epl2} 
\usepackage{bm}
\newcommand{\be}{\begin{equation}}
\newcommand{\bea}{\begin{eqnarray}}
\newcommand{\ee}{\end{equation}}
\newcommand{\eea}{\end{eqnarray}}

\newcommand{\cur}{\mbox{\scriptsize cur}}
\newcommand{\con}{\mbox{\scriptsize con}}
\newcommand{\dyn}{\mbox{\scriptsize dyn}}
\newcommand{\CL}{\mbox{\scriptsize CL}}

\newcommand{\mhcur}{\hat{m}_{\mbox{\scriptsize cur}}}

\newcommand{\mdyn}{m_{\mbox{\scriptsize dyn}}}

\newcommand{\mhcon}{\hat{m}_{\mbox{\scriptsize con}}}
\newcommand{\mscon}{m_{s,\mbox{\scriptsize con}}}

\newcommand{\qllsm}{QLL$\sigma$M}
\title{Constituent and current quark masses at low chiral energies}
\shorttitle{Constituent and current quark masses at low chiral energies}

\author{M.~D.~Scadron\inst{1,2} \and F.~Kleefeld\inst{2} \and G.~Rupp\inst{2}}
\shortauthor{M.~D.~Scadron \etal}
 
\institute{                    
  \inst{1} Physics Department, University of Arizona, Tucson, AZ 85721, USA \\
  \inst{2} Centro de F\'{\i}sica das Interac\c{c}\~{o}es Fundamentais,
           Instituto Superior T\'{e}cnico, Universidade T\'{e}cnica de Lisboa, 
           \\ P-1049-001 Lisboa, Portugal
}
\pacs{14.65.Bt}{Light quarks}
\pacs{14.40.Aq}{pi, K, and eta mesons}
\pacs{11.30.Rd}{Chiral symmetries}

\abstract{
Light constituent quark masses and the corresponding dynamical quark masses
are determined by data, the Quark-Level Linear $\sigma$ Model, and infrared
QCD. This allows to define effective nonstrange and strange current quark
masses which reproduce the experimental pion and kaon masses very accurately,
by simple additivity. Moreover, the masses of the light scalar mesons
$\sigma(600)$ and $\kappa(800)$ can be obtained straightforwardly from the
constituent quark masses. In contrast, the usual nonstrange and strange current
quark masses employed by Chiral Perturbation Theory do not allow a simple
quantitative explanation of the pion and kaon masses.  
}

\begin{document}

\maketitle

\section{Introduction}
In the chiral limit (CL), both the kaon and pion masses vanish. But away
from the CL, the chiral-symmetry breaking (ChSB) pion and kaon masses are
measured as \cite{PDG06} $m_{\pi^\pm}=139.57$ MeV, $m_{\pi^0}=134.98$ MeV,
$m_{K^\pm}=493.677$ MeV and $m_{K^0}=497.648$ MeV (dropping the small
experimental errors). The kaon $\bar{q}q$ mass difference is in fact
\cite{SKR07,SDR06}
\bea
\lefteqn{m_{K^0}(\bar{s}d)-m_{K^+}(\bar{s}u)\;=\;(m_d-m_u)_{\cur}=}\nonumber
\\[1mm] && (497.648-493.677)\; \mbox{MeV} \; = \; 3.97\;\mbox{MeV} \; ,
\label{kaonmd}
\eea
where the strange-current-quark-mass component cancels out. Likewise, the
difference of baryon masses composed of $qqq$ constituent quarks for
$\Sigma^-$ and $\Sigma^+$ is \cite{PDG06,SKR07}
\bea
\lefteqn{m_{\Sigma^-(sdd)} - m_{\Sigma^+(suu)} \; = \; 8.08\;\mbox{MeV}}
\nonumber \\[1mm] && \Rightarrow (m_d-m_u)_{\con} \; = \; 4.04\;\mbox{MeV}\;,
\label{Sigmamd}
\eea
where the strange-constituent-quark-mass component also cancels out. 

Since the strong-interaction bulk dynamical quark mass $m_{\dyn}=m_{\con}-
m_{\cur}$ conserves isospin, it should not be surprising that at low
energies both the constituent and current quark-mass differences are
\cite{SDR06}
\be
(m_d-m_u)_{\con} \; = \; (m_d-m_u)_{\cur} \; \approx \; 4 \; \mbox{MeV} \; ,
\label{curcon}
\ee
compatible with Eqs.~(\ref{kaonmd},\ref{Sigmamd}) above.
Moreover, a bulk dynamical CL nonstrange quark mass can be estimated as
\be
\mdyn \; \approx \; \frac{m_N}{3} \; \approx \; 313 \; \mbox{MeV} \; ,
\label{mdynN}
\ee
while the CL pion charge radius $r_\pi^{\CL}\approx0.63$ fm (as found from
vector-meson dominance \cite{S60}) implies
\be
\mdyn \; \approx \; \frac{\hbar c}{r_\pi^{\CL}} \; = \;
\frac{197.3\;\mbox{MeV$\cdot$fm}}{0.63\;\mbox{fm}}\; = \;313\;\mbox{MeV}\;.
\label{mdynr}
\ee
Lastly, low-energy QCD requires
\be
m_{\dyn} \; = \; \left[\frac{4\pi}{3}\,\alpha_s\,\langle-\bar{q}q\rangle
\right]^{\frac{1}{3}} \; .
\label{mdynqcd}
\ee
Then, demanding the latter to equal again 313 MeV in turn suggests
\cite{SKR06,ES84}
\be
\alpha_s(\mbox{1 GeV}) \; \approx \; 0.50 \;\;\; , \;\;\;
\langle-\bar{q}q\rangle^{\frac{1}{3}} \; \approx \; 245 \; \mbox{MeV} \; .
\label{alphaqq}
\ee

This further suggests \cite{ES84} a meson-quark coupling, for $N_c=3$,
\be
g_{\pi qq} \; = \; \frac{2\pi}{\sqrt{3}} \;\;\; , \;\;\;
\alpha_s^{\mbox{\scriptsize eff}} \; = \; \frac{g^2}{4\pi} \; = \;
\frac{\pi}{3} \; \sim \; 1 \; ,
\label{alphag}
\ee
so that the scalar $\sigma$ meson resonance condenses at a value twice
that of the average constituent quark mass given by the nucleon magnetic
moment in the nonrelativistic quark model (NRQM), viz.\ \cite{SDR06}
\be
\hat{m}_{\con} \; = \; \frac{1}{2}(m_u+m_d)_{\con} \; \approx \; 337.5
\; \mbox{MeV} \; .
\label{mudcon}
\ee
This occurs at a scalar mass $m_\sigma\approx$ 630--660 MeV. Note that
Eq.~(\ref{mudcon}) is very near the constituent mass resulting from the
quark-model chiral Goldberger--Treiman relation \cite{SDR06,W90a}, i.e.,
\be
\hat{m}_{\con} \; = \; f_\pi\,g_{\pi qq} \; \approx \; 93\:\mbox{MeV}
\times\frac{2\pi}{\sqrt{3}} \; \approx \; 337.4\;\mbox{MeV} \; ,
\label{mcongtr}
\ee
with 
\be
m_\sigma \; = \; 2m_{\dyn} \; \approx \; 630 \; \mbox{MeV} 
\label{msigma}
\ee
in the CL.

\section{Effective current quark mass in QCD}
Away from the CL, the effective current quark mass is
\be
\hat{m}_{\cur} \; = \; \hat{m}_{\con} - m_{\dyn} \; ,
\label{mcur}
\ee
where $m_{\dyn}$ in QCD runs as
\be
m_{\dyn}(p^2) \; \propto \; p^{-2} \; .
\label{mdynrun}
\ee
On the $\hat{m}_{\con}\approx337.5$ MeV mass shell, self-consistency then
requires
\be
m_{\dyn}(p^2\!=\!\hat{m}^2)=\frac{m_{\dyn}^3}{\hat{m}^2}=
\frac{(313)^3}{(337.5)^2}\;\mbox{MeV}=269.2\;\mbox{MeV} \; ,
\label{mdynself}
\ee
generating an effective current quark mass, via Eq.~(\ref{mcur}),
\be
\hat{m}_{\cur}^{\mbox{\scriptsize eff}} \; = \; (337.5-269.2)\;\mbox{MeV}
\; = \; 68.3 \; \mbox{MeV} \; .
\label{mcureff}
\ee
The latter is very near the ChSB pion-nucleon sigma term, from different
analyses:
\be
\left\{\begin{array}{l}
\sigma_{\pi N} \; = \; (55\pm13)\;\mbox{MeV \cite{HJS71}} \; , \\
\sigma_{\pi N} \; = \; (66\pm 9)\,\;\;\mbox{MeV \cite{NO74}} \; , \\
\sigma_{\pi N} \; = \; (64\pm 8)\,\;\;\mbox{MeV \cite{K82}} \; . 
\label{sigmaterm}
\end{array}\right.
\ee

\section{Pion \boldmath{$\bar{q}q$} mass}
Given $\hat{m}_{\cur}^{\mbox{\scriptsize eff}}$ from Eq.~(\ref{mcureff}),
the ChSB $\bar{q}q$ pion mass is \cite{SKR07,SKR06}
\be
\overline{m}_{\pi} \; = \; 2\hat{m}_{\cur}^{\mbox{\scriptsize eff}} \;=\;
136.6\;\mbox{MeV} \; ,
\label{mpiav}
\ee
almost midway between the observed \cite{PDG06} $m_{\pi^0}=134.98$ MeV
and $m_{\pi^\pm}=139.57$ MeV. In fact, the ChSB isotriplet pion masses
$m_{\pi^\pm}=\overline{m}_\pi+2\varepsilon$ and $m_{\pi^0}=
\overline{m}_\pi-\varepsilon$ have average value
\be
\overline{m}_\pi \; = \; \frac{1}{3}(m_{\pi^\pm}+2m_{\pi^0}) \; = \; 
136.5 \; \mbox{MeV} \; ,
\label{mpiavex}
\ee
very near the above estimate of 136.6 MeV.

\section{Kaon \boldmath{$\bar{q}q$} masses}
Given $m_d-m_u\approx4$ MeV and neglecting small experimental errors,
the observed ChSB kaon masses are
\begin{eqnarray}
\lefteqn{m_{K^+(\bar{s}u)}\;=\;m_{s,\cur}+m_{u,\cur} = } \nonumber \\[1mm]
&\hat{m}_{\cur}\left[1+\left(\frac{m_s}{\hat{m}}\right)_{\!\cur}\right]
-2\;\mbox{MeV} \; = \; 493.677\;\mbox{MeV} \; , \nonumber \\[1mm]
\lefteqn{m_{K^0(\bar{s}d)} \;=\; m_{s,\cur}+m_{d,\cur} = }  \\[1mm]
&\hat{m}_{\cur}\left[1+\left(\frac{m_s}{\hat{m}}\right)_{\!\cur}\right]
+2\;\mbox{MeV} \;=\; 497.648\;\mbox{MeV} \; . \nonumber
\label{mkpkz}
\end{eqnarray}
For $\hat{m}_{\cur}^{\mbox{\scriptsize eff}}$, this gives in {\em both}
\/cases, for $m_\pi=2\hat{m}_{\cur}$,
\be
\left(\frac{m_s}{\hat{m}}\right)_{\cur} \; = \; 
\frac{2\times 495.7\;\mbox{MeV}}{136.6\;\mbox{MeV}} \, - 1
\; \approx \; 6.26 \; , 
\label{msmhcur}
\ee
taking the average kaon mass as $(m_{K^+}+m_{K^0})/2=495.7$ MeV. The latter
ratio compares well with the light-plane \cite{SS75} and $\pi\pi$ \cite{SSF93}
predictions 6--7 and 6.33, respectively.

Independently of the above ratios 6.26, 6--7, 6.33, the ``good-bad'' chiral
operators lead to the infinite-momentum-frame (IMF) estimate \cite{GMS76}
\be
\left(\frac{m_s}{\hat{m}}\right)_{\cur}^{\mbox{\scriptsize IMF}} \; = \;
\left[\left(\frac{2m_K^2}{m_\pi^2}\right)-1\right]^{\frac{1}{2}}\;\approx\;5\;,
\label{msmhatimf}
\ee
requiring both squared meson and current quark masses in the IMF.

\section{\boldmath{$\left(m_s/\hat{m}\right)_{\cur}$}
from nonrenormalisation theorem}
Dealing instead with chiral current algebra, the vector-current divergence
$i\partial V^{6+i7}$ generates \cite{SF89a}
\be
\sqrt{2}\langle\pi^0|i\partial V^{6+i7}|\bar{K}^0\rangle \; = \; 
f_+(0)(m_K^2-m_\pi^2) \; ,
\label{nrth}
\ee
for the nonrenormalisation-theorem \cite{AG64} value \cite{PS84}
\be
f_+(0) \; = \; 1-\frac{g^2_{\pi qq}\delta^2}{8\pi^2} \; \approx \;
0.9677 \; ,
\label{fpz}
\ee
for $\delta=(m_s/\hat{m})_{\con}-1\approx0.44$. Then, with $f_K/f_\pi\approx
1.22$ \cite{PDG06} and $f_K/\left(f_\pi f_+(0)\right)\equiv X\approx1.2607$,
Eqs.~(\ref{nrth},\ref{fpz}) generate the current-quark-mass ratio \cite{SF89a}
(for $\overline{m}_\pi^2/\overline{m}_K^2\approx0.07594$)
\be
\left(\frac{m_s}{\hat{m}}\right)_{\cur} \; = \;
\frac{X+1-\frac{\overline{m}_\pi^2}{\overline{m}_K^2}}
     {X-1+\frac{\overline{m}_\pi^2}{\overline{m}_K^2}}\approx6.49\;.
\label{msmhatnrth}
\ee
The latter ratio is quite near the values 6--7, 6.26, 6.33 and 5 found
above. Note that, as $X\to1$,
$(m_s/\hat{m})_{\cur}\to2\overline{m}_K^2/\overline{m}_\pi^2-1$, that is,
the canonical chiral-perturbation-theory (ChPT) value (see
Eq.~(\ref{msmhatchpt}) below).

The current-divergence procedure leading to the ratio in Eq.~(\ref{msmhatnrth})
can be extended to the constituent-quark decuplet-baryon mass difference and
current-quark mass difference in $\Xi^-(ssd)-\Xi^0(ssu)$, using $SU(6)$
$d/f$ ratios \cite{SF89b}. In any case, a slightly larger $\sigma$-resonance
mass (650--670 MeV) is obtained with standard dispersion theory and
quark-level-linear-$\sigma$-model (\qllsm) \cite{DS95} techniques, but
\em not \em \/using 1- or 2-loop ChPT \cite{GL82,GLS91,L92}.

\section{Pion-nucleon \boldmath{$\sigma$}-term}
Specifically, ChPT before 1982 \cite{GL82} suggested that the $\pi N$
$\sigma$-term was near the Gell-Mann--Oakes--Renner (GMOR) \cite{GMOR68}
value of 26 MeV, and then the observed $\sigma$ resonance was unrelated to
the original ChPT (see Ref.~\cite{GL84}, App.\ B). However, in 1991 ChPT
instead claimed a $\pi N$ $\sigma$-term
of 60 MeV follows from the positive and coherent sum of \em four \em \/terms
\cite{GLS91,L92}, at the Cheng--Dashen (CD) \cite{CD71} point $t=2m_\pi^2$
(in MeV), i.e.,
\begin{eqnarray}
\sigma_{\pi N}(t=2m^2_\pi) & = & \sigma_{\pi N}^{\mbox{\scriptsize GMOR}} +
\sigma_{\pi N}^{\mbox{\scriptsize HOChPT}} + \sigma_{\pi N}^{\bar{s}s} +
\sigma_{\pi N}^{t\mbox{-\scriptsize dep.}} \nonumber \\
& \approx & (25 + 10 + 10 + 15)\;\mbox{MeV}  =  60 \; \mbox{MeV} \; .
\nonumber \\
\label{sigmachpt}
\end{eqnarray}
Here, the second term on the right-hand side arises from higher-order ChPT,
the third one from the strange-quark sea, and the fourth is a $t$-dependent
contribution due to going from $t=0$ to the CD point
$t=2m^2_\pi$, where the $\pi N$ background is minimal. Leutwyler \cite{L92}
concluded: \em ``The three pieces happen to have the same sign.'' \em Of
course, for things to work out right, all \em four \em \/pieces must have
the same sign, including the GMOR term, which reads
\be
\sigma_{\pi N}^{\mbox{\scriptsize GMOR}} \; = \; 
\frac{m_\Xi+m_\Sigma-2m_N}{2}\left(\frac{m_\pi^2}{m_K^2-m_\pi^2}\right)
\; \approx \; 26 \; \mbox{MeV} \; .
\label{sigmagmor}
\ee
We prefer instead to invoke the model-independent IMF version, which has
only \em one \em \/net term, viz.\
\be
\sigma_{\pi N}^{\mbox{\scriptsize IMF}} \; = \;
\frac{m^2_\Xi+m^2_\Sigma-2m^2_N}{2m_N}\left(\frac{m_\pi^2}{m_K^2-m_\pi^2}
\right) \; \approx \; 63 \; \mbox{MeV} \; ,
\label{sigmaimf}
\ee
which is compatible with observation \cite{HJS71,NO74,K82}, and also with
the ChSB effective current quark mass of about 68 MeV as found in
Eq.~(\ref{mcureff}) above.

\section{Strange current and constituent quark masses}
Given the average kaon mass and Eq.~(\ref{kaonmd}) --- but now with a plus
sign --- we get for a $\bar{q}q$ kaon
\be
\begin{array}{r}
\overline{m}_K  =  \displaystyle\frac{1}{2}\,
[m_{K^0}(\bar{s}d)+m_{K^+}(\bar{s}u)]  \approx  495.7\;\mbox{MeV} \:\;\;
\\[1mm]  =  (m_s+\hat{m})_{\cur} \; .
\label{kaonavm}
\end{array}
\ee
Subtracting now from Eq.~(\ref{kaonavm}) the nonstrange current quark mass 
$\hat{m}_{\cur}^{\mbox{\scriptsize eff}}=68.3$ MeV derived in
Eq.~(\ref{mcureff}), we deduce a strange current quark mass of
\be
m_{s,\cur} \; = \; (495.7-68.3)\;\mbox{MeV} \; = \; 427.4\;\mbox{MeV}\; ,
\label{mscur}
\ee
which is of course very near the current-quark-ratio version
\be
m_{s,\cur} = \left(\frac{m_s}{\hat{m}}\right)_{\!\cur}\!\!\times\mhcur
\approx 6.26\times68.3\;\mbox{MeV} = 427.5\;\mbox{MeV}\; .
\label{mscurratio}
\ee
As for the strange constituent quark mass, we may obtain an estimate
using the chiral quark-level Goldberger--Treiman relation
for \em constituent \em \/quarks:
\bea
\lefteqn{\left.\begin{array}{l}\displaystyle
\frac{1}{2}(m_s+\hat{m})_{\con} = f_K\,g_{Kqq} \;,\;\; g_{Kqq} = g_{\pi qq}
\; \mbox{\cite{SDR06}, \,\,or}  \\[3mm] \displaystyle
\left(\frac{m_s+\hat{m}}{2\hat{m}}\right)_{\con} \; = \;
\frac{f_K}{f_\pi} \; \approx \; 1.22 \; \mbox{\cite{PDG06}}
\end{array} \right\}\Rightarrow} \nonumber \\[2mm]
& m_{s,\con} = 1.44\,\mhcon = 1.44\times337.5\;\mbox{MeV}
\approx 486\;\mbox{MeV} \; , \!\!\!\! \nonumber \\ &
\label{mscon}
\eea
near the strange-quark valence value of 515 MeV \cite{BLP64}.

Alternatively, let us estimate $\mscon$ from the $\Sigma$ baryon mass
as
\be
\frac{m_{\Sigma^-}+m_{\Sigma +}}{2} \; = \; 1193\;\mbox{MeV}
\; = \; \mscon+2\mhcon \; ,
\label{msconsigma}
\ee
which yields $\mscon\approx518$ MeV, close to the values 515 MeV and
486 MeV above. A similar constituent strange quark mass also follows
from the vector-meson $\bar{s}s$ $\phi$(1020) mass, suggesting
$\mscon \approx 1020/2$ MeV $=510$ MeV.
%Applying now the analogue of Eq.~(\ref{mdynr}) to the kaon charge radius
%$r_K=0.56$ fm \cite{PDG06}, we get
%\be
%m_{s,\dyn}\;\sim\;\frac{\hbar c}{r_K}\;\approx\;
%\frac{197.3\;\mbox{MeV$\cdot$fm}}{0.56\;\mbox{fm}}\;
%\approx\;352\;\mbox{MeV} \; .
%\label{mdynr}
%\ee
%Similarly to Eq.~(\ref{mdynqcd}), this gives for the dynamical strange
%quark mass, on the $m_{s,\con}=486$ MeV mass shell,
%\be
%m_{s,\dyn}(p^2=m^2_{s,\con}) \; = \; \frac{m^3_{s,\dyn}}{m^2_{s,\con}}
%\; \approx \; 185 \; \mbox{MeV},
%\label{mdynrun}
%\ee
%which results in a strange current quark mass
%\be
%m_{s,\cur} \; = \; (486 - 185) \; \mbox{MeV} \; \sim \; 300\;\mbox{MeV} \;.
%\label{mscur}
%\ee

\section{\boldmath{$\left(m_s/\hat{m}\right)_{\cur}$} in ChPT}

Alternatively, we may compare this with the ChPT prediction for the
current quark mass (see e.g.\ review from 1982 \cite{GL82}), for
$f_\pi\approx93$ MeV,
\be
\mhcur \; = \; \frac{(f_\pi m_\pi)^2}{2\langle-\bar{q}q\rangle}
\; \approx \; 5.6\;\mbox{MeV} \; ,
\label{mcurchpt}
\ee
again using $\langle-\bar{q}q\rangle^{1/3}\approx245$ MeV. When this is
combined with the ChPT ratio \cite{GL82}
\be
\left(\frac{m_s}{\hat{m}}\right)_{\cur} \; \approx \;
\frac{2m_K^2}{m_\pi^2}-1 \; \approx \; 25.3 \; ,
\label{msmhatchpt}
\ee
one could infer a strange current quark mass of 
\be
m_{s,\cur}\;\approx\;25.3\times5.6\;\mbox{MeV}\;\approx\;142\;\mbox{MeV}\;,
\label{mscurchpt}
\ee
so only about 29\% of the strange constituent quark mass of 486 MeV
found in Eq.~(\ref{mscon}), which does not seem very realistic. This is to
be contrasted with $m_{s,\cur}/m_{s,\con}\approx88$\% from our
Eq.~(\ref{mscur}) above.

Studying the ChPT Eqs.~(\ref{mcurchpt},\ref{msmhatchpt}) independently of one
another, the ChSB nonstrange relation in Eq.~(\ref{mcurchpt}) appears to be
unrelated to the observed ChSB $\pi N$ $\sigma$-terms in
Refs.~\cite{HJS71,NO74,K82}, and also to the observed $\bar{q}q$ pion mass
of 138 MeV, which we managed to simply relate to the effective nonstrange
current quark mass in Eqs.~(\ref{mcureff},\ref{mpiav}) as
$m_\pi\approx2m_{\cur}^{\mbox{\scriptsize eff}}$. Moreover, the ratio in
Eq.~(\ref{msmhatchpt}) is about a factor 4 larger than the values 6.26
from the average kaon mass in Eq.~(\ref{msmhcur}), 6--7 from the
light plane \cite{SS75}, 6.33 from $\pi\pi$ scattering \cite{SSF93}, and
6.49 from the current-divergence nonrenormalisation
relation~(\ref{msmhatnrth}).
 
Finally, we refer to an  unpublished paper from 1974 \cite{FGML74} also
finding a current-quark-mass ratio $(m_s/\hat{m})_{\cur}\sim6$, which
avoided ``bad'' quark operators such as $\bar{q}\lambda_q\gamma_5 q$
in Ref.~\cite{GMOR68}. Note that in the latter reference, GMOR suggested
$f_K=f_\pi$ to first order away from the CL. We argue that the ChSB
theory as employed in Refs.~\cite{GL82,GLS91,L92,CCL06} must invoke
a 22\% ChSB increase even in first order, because data says \cite{PDG06}
$f_K/f_\pi\approx1.22$, \em not \em \/$f_K\approx f_\pi$.

With hindsight, Ref.~\cite{FGML74} appears to realise that the standard
perturbative Taylor-series approach to the ratio of strange over nonstrange
current quark mass \em breaks down \em \/ when $f_K\to f_\pi$, as we have
shown in Eq.~(\ref{msmhatnrth}), which does \em not \em \/give
Eq.~(\ref{msmhatchpt}) as ChPT suggests. Even \em generalised \em \/ChPT
admits the possibility that this ratio may be considerably smaller
\cite{DGS00} than the standard ChPT value $\sim25$.

We are well aware that recent lattice analyses report values for the
ratio $m_s/\hat{m}$ close to the ChPT prediction, namely 25.7
(CP-PACS and JLQCD Collaborations) \cite{I07}, 27.2 (HPQCD Collaboration)
\cite{M06}, 25.3 (QCDSF Collaboration) \cite{G04} and 27.4 (MILC, HPQCD and
UKQCD Collaborations) \cite{A04}. However, this is not so surprising, as
all these lattice results, obtained at a scale of 2 GeV in the
$\overline{\mbox{MS}}$ scheme, strongly rely upon perturbative chiral
extrapolations in the framework of \em standard \em \/ChPT. Such
extrapolations assume the validity of the GMOR relation, ruling out significant
corrections to it, in contrast with approaches like \em generalized \em \/ChPT
and also the \qllsm. Let us furthermore quote from the most recent of these
lattice works, namely Ref.~\cite{I07}: \em ``We note that our WChPT fits to
data do not exhibit a clear chiral logarithm, probably because $u$ and $d$
quark masses in our simulation are not sufficiently small.'' \em In view of
these and other uncertainties, like finite-volume effects, we believe there is
no conflict between the very small current quark masses found in such
lattice/ChPT approaches and the \em effective \em \/current quark masses in the
present paper.

\section{Scalar-meson masses and summary}
With regard to the nonstrange and strange constituent quark masses, we may
estimate the corresponding scalar-meson masses $m_\sigma$ and $m_\kappa$,
where $\sigma$ and $\kappa$ stand for the PDG states $f_0(600)$ and
$K_0^*(800)$ \cite{PDG06}, respectively. For the $\sigma$, we get \cite{SKR06}
\be
m_\sigma\;\approx\;2\hat{m}_{\con}\;\approx\;
625\,\mbox{--}\,675 \;\mbox{MeV}\;,
\label{msigmaclncl}
\ee
depedending on whether we work in the CL for $\hat{m}_{\con}$ or not. Regarding
the $\kappa$, the resulting mass range is \cite{SKR06}
\be
m_\kappa\;\approx\;2\sqrt{m_{s,\con}\,\hat{m}_{\con}}\;\approx\;
780\,\mbox{--}\,810 \;\mbox{MeV}\;.
\label{mkappaclncl}
\ee
The latter range is entirely compatible with the E791 \cite{A02} value
\be
m_\kappa^{\mbox{\scriptsize E791}} \; = \; (791\pm19)\;\mbox{MeV}\;,
\label{mkappae791}
\ee
found from $D^+\to K^-\pi^+\pi^+$ decay, and also with the resonance peak
masses obtained in the unitarised quark-meson models of Ref.~\cite{B86}.

ChPT, on the other hand, has a much more complicated relationship with 
the light scalar mesons. During many years, ChPT disputed the relevance
or even the very existence  of the $\sigma$ meson. However, recently
\cite{CCL06} Roy equations were employed to generate a $\sigma$ pole, which
indeed is in rough agreement with Weinberg's \cite{W90b} estimate
$\Gamma_\sigma\approx9\Gamma_\rho/2$. These equations amount to
twice-subtracted dispersion relations in which the two subtraction
constants are fixed via higher-order ChPT. While such a unitarisation of ChPT
``by hand'' is not really new (see e.g.\ Ref.~\cite{OOP99}), completely
unprecedented is the claim to have pinned down the $\sigma$ pole postition
with surprisingly small error bars. However, the latter cannot be trusted
in view of the neglected $\sigma$-$f_0(980)$ mixing and the deficient
treatment of the $\bar{K}K$ channel \cite{B06}. Moreover, the Roy-equation
approach as applied in Ref.~\cite{CCL06} may be at odds \cite{K07} with 
the standard Chew-Mandelstam \cite{CM60} double-dispersion representation
of $\pi\pi$ amplitudes \cite{DS95}. For all that, even the scattering lengths
entering the Roy equations are to be questioned, as they are related to scalar
radii which yield ChSB effects of the order of 6--8\%, instead of the
observed 3\%.

In summary, the present paper shows how constituent, dynamical and current
quark masses --- compatible with QCD expectations --- can be defined
that are related in a very simple way to light meson masses and other
low-energy observables. In particular, the thus determined effective
nonstrange and strange current quark masses determine the pion and kaon
masses just by additivity. Moreover, the latter nonstrange current quark
mass is of the same size as another crucial measure of ChSB, viz.\
the $\pi N$ $\sigma$-term. Finally, our constituent quark masses allow
to straighforwardly obtain the light scalar meson masses $m_\sigma$ and
$m_\kappa$ in agreement with experiment. In contrast, ChPT does not allow
any simple relation between quark masses and light-meson masses or the
$\pi N$ $\sigma$-term.

\acknowledgments
This work was partially supported by the {\it Funda\c{c}\~{a}o para a
Ci\^{e}ncia e a Tecnologia} \/of the {\it Minist\'{e}rio da Ci\^{e}ncia,
Tecnologia e Ensino Superior} \/of Portugal, under contract PDCT/FP/63907/2005.

\end{document}